\documentclass[aps,prd,twocolumn,superscriptaddress,nofootinbib]{revtex4-1}

\usepackage{latexsym}
\usepackage{amsmath}
\usepackage{amssymb}
\usepackage{amsfonts}
\usepackage{bm}
\RequirePackage{lineno}

\usepackage{color}
\definecolor{purple}{rgb}{0.5,0,0.5}
\definecolor{blue}{rgb}{0.0,0,0.9}
\definecolor{prdblue}{rgb}{0.133,0.118,0.498}
\usepackage[colorlinks=true, pdfstartview=FitV, linkcolor=prdblue, citecolor= prdblue, urlcolor=prdblue]{hyperref}

\usepackage{supertabular} 
\usepackage{placeins}
\usepackage{epsfig}
\usepackage{graphicx}
\usepackage{hyperref}

\hypersetup{
  breaklinks=true,
  colorlinks = true,
  linkcolor = blue,
  anchorcolor = blue,
  citecolor = blue,
  filecolor = blue,
  pagecolor = blue,
  urlcolor = blue
}
\hypersetup{
  bookmarks=true,
  bookmarksnumbered=true,
  bookmarkstype=toc,
  linktocpage=true
}

\begin{document}

\modulolinenumbers[2]

\setlength{\oddsidemargin}{-0.5cm} \addtolength{\topmargin}{15mm}

\title{\boldmath Improved measurement of the absolute branching fraction of inclusive semileptonic $\Lambda_c^+$ decay}
\author{
  \small  
M.~Ablikim$^{1}$, M.~N.~Achasov$^{11,b}$, P.~Adlarson$^{70}$, M.~Albrecht$^{4}$, R.~Aliberti$^{31}$, A.~Amoroso$^{69A,69C}$, M.~R.~An$^{35}$, Q.~An$^{66,53}$, X.~H.~Bai$^{61}$, Y.~Bai$^{52}$
% ...
  \\
   \vspace{0.2cm}
   (BESIII Collaboration)\\
   \vspace{0.2cm} {\it
$^{1}$ Institute of High Energy Physics, Beijing 100049, People's Republic of China\\
% ...
$^{4}$ Bochum Ruhr-University, D-44780 Bochum, Germany\\
% ...
$^{11}$ G.I. Budker Institute of Nuclear Physics SB RAS (BINP), Novosibirsk 630090, Russia\\
% ...
$^{31}$ Johannes Gutenberg University of Mainz, Johann-Joachim-Becher-Weg 45, D-55099 Mainz, Germany\\
% ...
$^{35}$ Liaoning Normal University, Dalian 116029, People's Republic of China\\
% ...
$^{52}$ Southeast University, Nanjing 211100, People's Republic of China\\
$^{53}$ State Key Laboratory of Particle Detection and Electronics, Beijing 100049, Hefei 230026, People's Republic of China\\
% ...
$^{61}$ University of Jinan, Jinan 250022, People's Republic of China\\
% ...
$^{66}$ University of Science and Technology of China, Hefei 230026, People's Republic of China\\
% ...
$^{69}$ University of Turin and INFN, (A)University of Turin, I-10125, Turin, Italy; (B)University of Eastern Piedmont, I-15121, Alessandria, Italy; (C)INFN, I-10125, Turin, Italy\\
$^{70}$ Uppsala University, Box 516, SE-75120 Uppsala, Sweden\\
% ...
$^{b}$ Also at the Novosibirsk State University, Novosibirsk, 630090, Russia\\
% ...
   \vspace{0.4cm}
}
}

\begin{abstract}
Using $4.5~\mathrm{fb}^{-1}$ of $e^+e^-$ annihilation data samples collected at center-of-mass energies ranging from 4.600 to 4.698~GeV with the BESIII detector at the BEPCII collider, we measured the absolute branching fraction for the inclusive semileptonic decay $\Lambda_c^+\rightarrow Xe^+\nu_e$, where $X$ refers to any possible particle system. The branching fraction of the decay is determined to be $\mathcal{B}(\Lambda^+_c\rightarrow Xe^+\nu_e)=(4.06\pm0.10_{\rm stat.}\pm0.09_{\rm syst.})\%$. 
Our result improves the precision of previous 
measurement of $\mathcal{B}(\Lambda^+_c\rightarrow Xe^+\nu_e)$ by more than threefold.
Using the known $\Lambda_c^+$ lifetime and the charge-averaged semileptonic decay width of nonstrange charmed mesons, we measure the ratio of inclusive semileptonic decay widths 
$\Gamma(\Lambda_c^+\rightarrow X e^+\nu_e)/\bar{\Gamma}(D\rightarrow Xe^+\nu_e)=1.28\pm0.05$, where statistical and systematic uncertainties are combined.
\end{abstract}


\maketitle

\section{Introduction}
The lowest-lying charmed baryon $\Lambda_c^+$ was discovered more than 40 years ago~\cite{PRL37_882,PRL44_10}, but about $30\%$ of $\Lambda_c^+$ decays are still unmeasured~\cite{pdg2020}.
In recent years, great experimental progress has been made in the study of $\Lambda_c^+$ semileptonic (SL) decays~\cite{CPC44_040001,NSR8_11}. 
For example, the BESIII experiment obtained a precise absolute branching fraction (BF) for the dominant semileptonic mode, $\mathcal{B}(\Lambda_c^+\rightarrow \Lambda e^+\nu_e)=(3.56\pm0.11_{\rm stat.}\pm0.07_{\rm syst.})\%$, and also measured the differential decay rate and form factors in this decay for the first time~\cite{Lamev}. 
The $\Lambda_c^+\rightarrow pK^- e^+\nu_e$ decay is now also observed with a significance of $8.2\sigma$ with a measured BF $\mathcal{B}(\Lambda_c^+\rightarrow pK^- e^+\nu_e)=(0.82\pm0.15_{\rm stat.}\pm0.06_{\rm syst.})\times 10^{-3}$~\cite{pKev}.  Evidence of the $\Lambda^*(1520)$ is seen in the $pK^-$ mass spectrum with a significance of $3.3\sigma$~\cite{pKev}.
In addition, the BF for the inclusive SL $\Lambda_c^+$ decay is measured to be $\mathcal{B}(\Lambda_c^+\rightarrow X e^+\nu_e)=(3.95\pm0.34_{\rm stat.}\pm0.09_{\rm syst.})\%$~\cite{PRL121_251801}. 
However, in comparison to the experimental studies of SL decays in $D$ mesons, measurements of the SL decays in $\Lambda_c^+$ are still very limited~\cite{pdg2020}. 

Improved precision on $\mathcal{B}(\Lambda_c^+\rightarrow X e^+\nu_e)$ 
is desirable.  
First, a comparison of $\mathcal{B}(\Lambda_c^+\rightarrow X e^+\nu_e)$ with $\mathcal{B}(\Lambda_c^+\rightarrow \Lambda e^+\nu_e)$ and $\mathcal{B}(\Lambda_c^+\rightarrow pK^- e^+\nu_e)$ indicates that some new $\Lambda_c^+$ SL decays may exist beyond the known $\Lambda_c^+\rightarrow \Lambda \ell^+\nu_{\ell}$ and $\Lambda_c^+\rightarrow pK^-\ell^+\nu_{\ell}$  ($\ell=e, \mu$). 
Second, combining this branching fraction with the known lifetime of the $\Lambda_c^+$ baryon determines the SL decay width $\Gamma(\Lambda_c^+\rightarrow Xe^+\nu_e)$. Comparing with the charge-averaged non-strange $D$ SL decay width $\bar{\Gamma}(D\rightarrow Xe^+\nu_e)$, one finds the ratio $\Gamma(\Lambda_c^+\rightarrow Xe^+\nu_e)/\bar{\Gamma}(D\rightarrow Xe^+\nu_e) = 1.26\pm0.12$~\cite{PRL121_251801}, about a $10\%$ uncertainty. 
Improving the precision on $\Gamma(\Lambda_c^+\rightarrow Xe^+\nu_e)/\bar{\Gamma}(D\rightarrow Xe^+\nu_e)$ is helpful for testing current theoretical predictions~\cite{PRD83_034025,PRD86_014017,PRD49_1310}.

This article presents an improved BF measurement of the $\Lambda_c^+$ inclusive SL decay. Our measurement is performed by using data sets collected with the BESIII detector at center-of-mass energies $\sqrt{s}=4.600, 4.612, 4.628, 4.640, 4.661, 4.682$ and $4.698$ GeV. The total integrated luminosity for these data samples is $4.5~\mathrm{fb}^{-1}$~\cite{lum_4600,lum_new}. These are the largest data samples collected near the $\Lambda_c^+\bar{\Lambda}_c^-$ production threshold. Throughout this paper, charge-conjugate modes are
implied unless explicitly noted.

\section{BESIII Detector and Monte Carlo Simulation}
The BESIII detector~\cite{Ablikim:2009aa} records symmetric $e^+e^-$collisions provided by the BEPCII storage ring in the center-of-mass energy range from 2.00 to 4.95~GeV, with a peak luminosity of $1 \times 10^{33}\;\text{cm}^{-2}\text{s}^{-1}$ achieved at $\sqrt{s} = 3.77\;\text{GeV}$. 
The BESIII spectrometer is a cylindrical detector with a solid-angle
coverage of 93\% of $4\pi$. The detector consists of a helium-gas
based main drift chamber (MDC), a plastic scintillator time-of-flight
(TOF) system, a CsI(Tl) electromagnetic calorimeter (EMC), a
superconducting solenoid providing a 1.0\,T magnetic field and a muon
counter~\cite{Muon}. The charged-particle momentum resolution is 0.5\% at a
transverse momentum of 1\,GeV$/c$, and the specific energy loss
(d$E$/d$x$) resolution is 6\% for the electrons from Bhabha scattering.
The photon energy resolution in the EMC is 2.5\% in the barrel and
5.0\% in the end-caps at energies of 1\,GeV.  The time resolution of
the TOF barrel part is 68 ps, while that of the end-cap part is 110
ps. The end-cap TOF system was upgraded in 2015 with multi-gap resistive plate chamber technology, providing a time resolution of 60 ps~\cite{tofup}.  
About 11\% of the dataset was obtained before the TOF upgrade.  
More details about the design and performance of the detector are
given in Ref.~\cite{Ablikim:2009aa}.

A {\sc geant4}-based~\cite{geant4} Monte Carlo (MC) simulation toolkit,
which includes the geometric description of the detector and the
detector response, is used to determine signal detection efficiency and
to estimate potential backgrounds. In the simulation, the effects of beam-energy spread and initial-state radiation~\cite{SJNP41_466} are incorporated using {\sc kkmc}~\cite{kkmc},
and final-state radiation from charged final state particles is incorporated using  {\sc photos}~\cite{plb303_163}.
An inclusive MC sample consisting of $\Lambda_c^+\bar{\Lambda}_c^-$, $D^{(*)}_{(s)}\bar{D}^{(*)}_{(s)}$ pairs, 
radiative return to charmonium(-like) $\psi$ states at lower masses and continuum processes of $q\bar{q}$ ($q=u,d,s$), along with Bhabha scattering, $\mu^+\mu^-$, $\tau^+\tau^-$ 
and $\gamma\gamma$ events are generated.
All particle decays are modelled with {\sc evtgen}~\cite{nima462_152} using BFs either taken from the Particle Data Group~\cite{pdg2020}, when available, or otherwise estimated with {\sc lundcharm}~\cite{lundcharm,lundcharm2}.

\section{Analysis}

The first procedure of this analysis is to reconstruct ``single-tag" (ST) events with a fully
reconstructed $\bar{\Lambda}_c^-$ candidate. The $\bar{\Lambda}_c^-$ hadronic decay tag modes used in this analysis are:
$\bar{\Lambda}^-_c\rightarrow \bar{p} K^0_S$, $\bar{p} K^+\pi^-$,
$\bar{p}K^0_S\pi^0$, $\bar{p} K^+\pi^-\pi^0$, $\bar{p}
K^0_S\pi^+\pi^-$, $\bar{\Lambda}\pi^-$, $\bar{\Lambda}\pi^-\pi^0$,
$\bar{\Lambda}\pi^-\pi^+\pi^-$, $\bar{\Sigma}^0\pi^-$,
$\bar{\Sigma}^-\pi^0$, $\bar{\Sigma}^-\pi^+\pi^-$ and $\bar{\Sigma}^0\pi^-\pi^0$, where the intermediate particles $K^0_S$, $\bar{\Lambda}$, $\bar{\Sigma}^0$,
$\bar{\Sigma}^-$ and $\pi^0$ are reconstructed via
$K^0_S\rightarrow \pi^+\pi^-$, $\bar{\Lambda}\rightarrow
\bar{p}\pi^+$, $\bar{\Sigma}^0\rightarrow \gamma\bar{\Lambda}$ with
$\bar{\Lambda}\rightarrow \bar{p}\pi^+$, $\bar{\Sigma}^-\rightarrow
\bar{p}\pi^0$ and $\pi^0\rightarrow \gamma\gamma$, respectively.
Starting with this ST sample, we identify the inclusive SL decay $\Lambda_c^+\rightarrow Xe^+\nu_e$ candidates; selected events are referred to as the double-tag (DT) sample~\cite{ST}.  

\subsection{Single-tag selection}

Charged tracks are required to satisfy 
$|\!\cos\theta|<0.93$, where $\theta$ is the polar angle with respect to to the $z$-axis, which is the symmetry axis of the MDC.
Their distances of the closest approach to the interaction point (IP) are required to be
less than 10\,cm along $z$-axis and less than 1\,cm in the
transverse plane. For charged tracks originating from $K^0_S$, $\bar{\Lambda}$ and $\bar{\Sigma}^-$, there is no transverse requirement and a relaxed requirement of 20\,cm matching along the $z$-axis is employed.  
The information from the d$E$/d$x$ in MDC and the flight time in the TOF are used
to obtain particle identification (PID) probabilities for the pion ($\mathcal{L}_{\pi}$) and
kaon ($\mathcal{L}_K$) hypotheses. The pion and kaon candidates are
selected using $\mathcal{L}_{\pi} > \mathcal{L}_{K}$ and
$\mathcal{L}_{K} > \mathcal{L}_{\pi}$, respectively. To identify protons, 
the information from the d$E$/d$x$, TOF, and EMC are combined
to calculate the PID probability $\mathcal{L'}$.  Charged tracks
satisfying $\mathcal{L'}_p> \mathcal{L'}_{\pi}$ and
$\mathcal{L'}_{p} > \mathcal{L'}_{K}$ are identified as proton
candidates.

Photon candidates are identified using showers in the EMC. 
The deposited energy of each shower must be more than 25\,MeV in the barrel region ($|\!\cos\theta| \le 0.80$) and more than 50\,MeV in the end cap region ($0.86 \le
|\!\cos\theta| \le 0.92$). 
To exclude showers that originate from charged tracks, the angle subtended by the EMC shower and the position of the closest charged track at the EMC must be greater than 10 degrees as measured from the IP. To suppress electronic noise and showers unrelated to the event, the difference between the EMC time and the event start time is required to be within [0, 700] ns. To reconstruct $\pi^0$
candidates, the invariant mass of the accepted photon pair is
required to be within $(0.110,~0.155)$~GeV$/c^2$. A kinematic fit is
performed to constrain the $\gamma\gamma$ invariant mass to the
known $\pi^0$ mass~\cite{pdg2020}, and the $\chi^2$ of the
kinematic fit is required to be less than 20. The fitted momenta of
the $\pi^0$ are used in the further analysis.

The $K^0_S$ meson is reconstructed from two oppositely charged tracks with invariant mass within $(0.485,~0.510)$~GeV$/c^2$.
The two charged tracks are constrained to originate from a common vertex and their decay length relative to the IP is required to be larger than zero.
 The invariant masses $M(\bar{p}\pi^+)$, $M(\gamma\bar{\Lambda})$ and $M(\bar{p}\pi^0)$ are required to be
within $(1.110,~1.121)$~GeV/$c^2$, $(1.179,~1.205)$~GeV/$c^2$ and $(1.171,~1.204)$~GeV/$c^2$ to reconstruct
candidates for $\bar{\Lambda}$, $\bar{\Sigma}^0$ and $\bar{\Sigma}^-$, respectively. 
For the ST mode $\bar{p}\pi^+\pi^-$, the backgrounds from Cabibbo-favored components of $\bar{\Lambda}\pi^-$ and
$\bar{p}K^0_S$ are removed by rejecting any event
with $M_{\bar{p}\pi^+}\in (1.105,1.125)$~GeV/$c^2$ and
$M_{\pi^+\pi^-}\in (0.475, 0.520)$~GeV/$c^2$.

\begin{figure}[htbp]
\begin{center}
   \begin{minipage}[t]{8.5cm}
   \includegraphics[width=\linewidth]{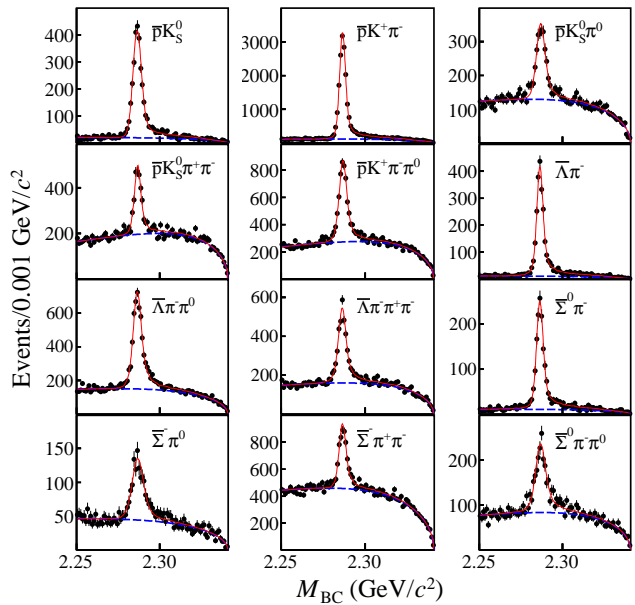}
   \end{minipage}
\caption{Fits to the $M_{\rm BC}$ distributions for different ST modes at $\sqrt{s}=4.682$~GeV.
The points with error bars are data, the (red) solid curves show the total
fits and the (blue) dashed curves are the fitted backgrounds. }
\label{fig:tag_lambdac}
\end{center}
\end{figure}

The ST $\bar{\Lambda}^-_c$ signals are identified using the beam
constrained mass: 
 $$M_{\rm BC}=\sqrt{(\sqrt{s}/2)^2-|\vec{p}_{\bar{\Lambda}^-_c}|^2},$$
where $\vec{p}_{\bar{\Lambda}^-_c}$ is the measured momentum of the ST $\bar{\Lambda}^-_c$.
The energy difference $\Delta E=\sqrt{s}/2-E_{\bar{\Lambda}^-_c}$ is determined for ST $\bar{\Lambda}^-_c$ baryons, where $E_{\bar{\Lambda}^-_c}$ is the measured energy of the ST $\bar{\Lambda}^-_c$.
If an event satisfies more than one $\bar{\Lambda}_c^-$ tag, only the tag with the minimum $|\Delta E|$ is kept to avoid double counting among STs (but a second tag with opposite charm is allowed). The $M_{\rm BC}$ distributions at $\sqrt{s}=4.682$~GeV for twelve ST modes are shown in Fig.~\ref{fig:tag_lambdac}. 
Unbinned maximum likelihood fits are performed to the spectra using the MC-simulated signal shape convolved with a Gaussian function accounting for
differences of resolutions between data and MC simulation to describe the signal, and an ARGUS function~\cite{plb241_278} with end-point fixed at $\sqrt{s}/2$ to describe the background.
The $\Delta E$ requirements, $M_{\rm BC}$ distributions for other data sets and their ST yields are documented in Refs.~\cite{Lamev,pKev}. 
The total ST yield reconstructed in all data samples is $N_{\rm ST}=115\,437\pm446$, where the uncertainty is statistical only.

\subsection{Double-tag selection}

In the selected ST sample of $\bar{\Lambda}_c^-$ candidates, we search for a charged track identified as a positron in the recoiling system of ST $\bar{\Lambda}_c^-$ baryons.
The charged track is required to have 
$|\!\cos\theta|<0.80$ to ensure that the track lies within the barrel of the EMC, which has better energy resolution than the EMC end-caps. The distance of the closest approach to the IP is required to be
less than 10\,cm along the $z$-axis and less than 1\,cm in the
transverse plane. 
The track is required to have momentum above 200 MeV/$c$ 
since PID is difficult at low momenta.  

We sort the selected recoil-side positron candidates into sixteen 50~MeV/$c$ momentum bins between 200~MeV/$c$ and 1000~MeV/$c$.
For a given momentum bin, the selected charged tracks are divided into two charge categories: the right-sign (RS) and wrong-sign (WS) samples, where the charge of the RS (WS) track is required to be opposite (equal) to that of the ST $\bar{\Lambda}_c^-$ candidate. 
Here, the WS samples are used to determine the charge symmetric backgrounds in the RS samples. 
For both charge categories in each of the momentum bins, the charged tracks are then sorted into four mutually exclusive PID hypotheses: positron, pion, kaon and proton. 
The PID of each selected charged track is implemented with the information measured from the d$E$/d$x$, TOF and EMC, and the combined likelihoods for positron, pion, kaon and proton hypotheses ($\mathcal{L'}_e$,
$\mathcal{L'}_\pi$, $\mathcal{L'}_K$ and $\mathcal{L'}_p$) are calculated.
The positron candidate must satisfy $\mathcal{L'}_{e} > 0.001$ and
$\mathcal{L'}_e/(\mathcal{L'}_e+\mathcal{L'}_{\pi}+\mathcal{L'}_K)>0.8$. To further suppress the background from charged hadrons, the $E_e/p_e>0.8$ is required~\cite{PRL121_251801,PRD104_012003}, where $E_e$ and $p_e$ are the deposited energy in the EMC and momentum value measured by the MDC, respectively. 
The remaining selected charged tracks are assigned to different hadron types according to the highest likelihood-value that is also greater than 0.001. 

\begin{figure}[tp!]
\begin{center}
   \begin{minipage}[t]{8cm}
   \includegraphics[width=\linewidth]{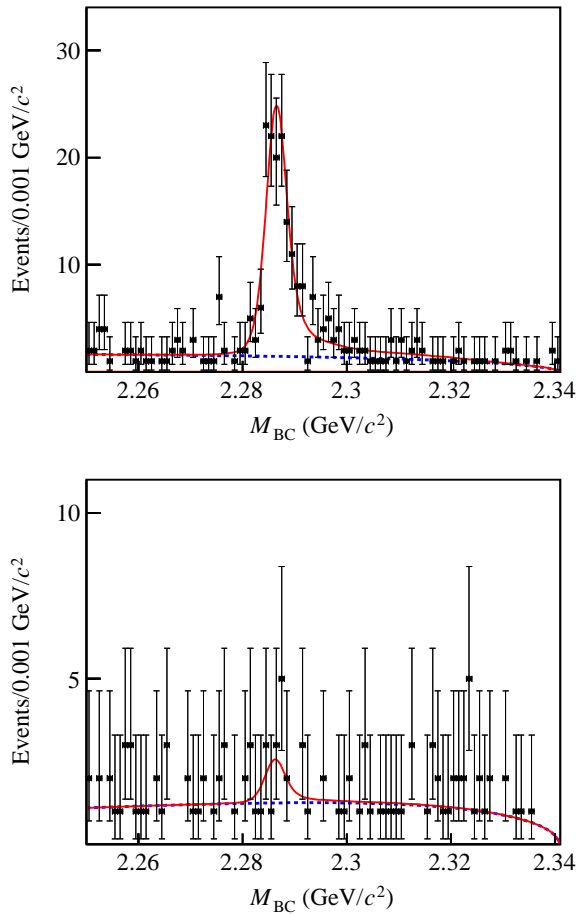}
   \end{minipage}
   \caption{ Example $M_{\rm BC}$ fits for double-tag RS (upper) and WS (lower) positron candidates with momentum in the range $500-550$~MeV/$c$ for data collected at $\sqrt{s}=4.682$~GeV. 
The points with error bars are data, the (red) solid curves show the total fits and the (blue) dashed curves are the fitted backgrounds. }
\label{fig:MbcFit}
\end{center}
\end{figure}

\begin{figure*}[tp!]
\begin{center}
   \begin{minipage}[t]{16cm}
   \includegraphics[width=\linewidth]{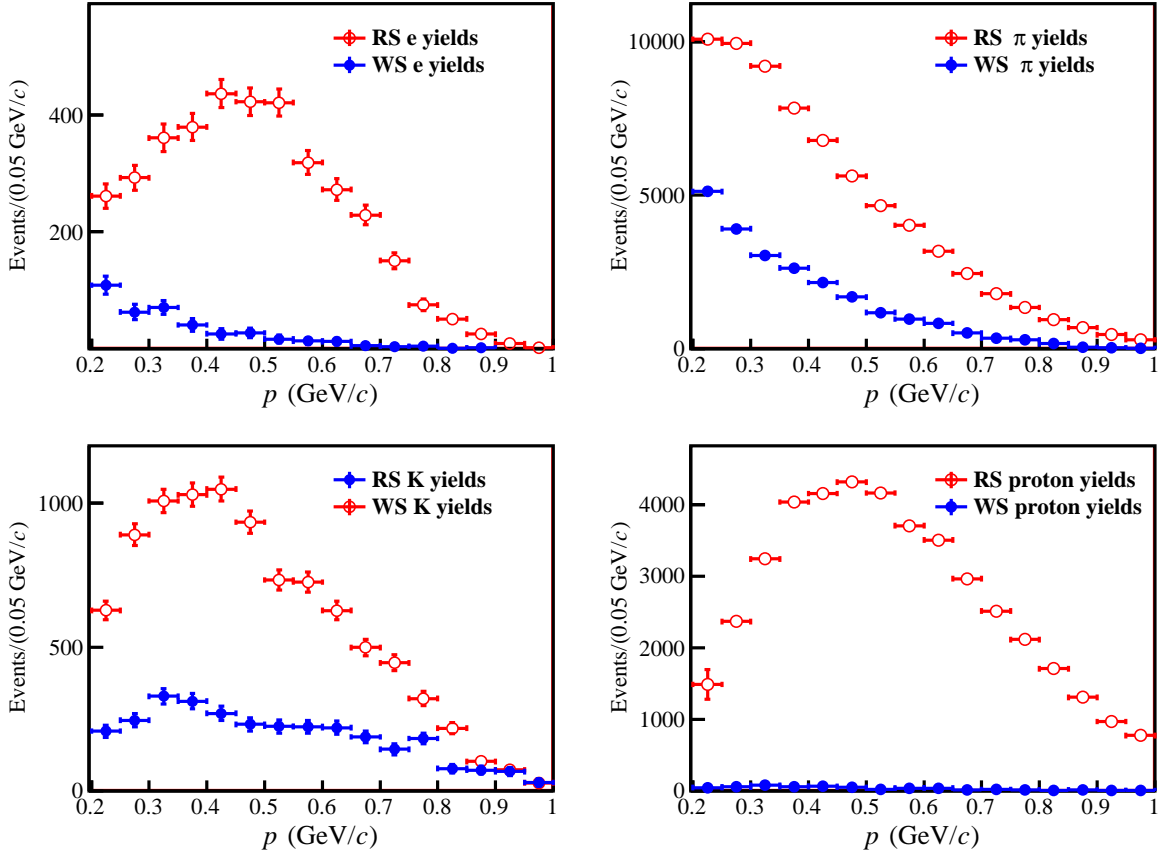}
   \end{minipage}
   \caption{Measured RS (blue) and WS (red) yields for each particle category as a function of momentum.}
\label{fig:Nobs}
\end{center}
\end{figure*}

For both RS and WS in a given momentum bin, we determine the number of the charged tracks originating from a $\Lambda_c^+$ by performing the unbinned fit to the $M_{\rm BC}$ distribution of the ST 
$\bar{\Lambda}_c^-$ candidate. Figure~\ref{fig:MbcFit} shows two examples of the fits, where the $M_{\rm BC}$ distributions obtained from all twelve $\bar{\Lambda}_c^-$ ST hadronic decays are combined together.
The reason for combining all ST modes together is that the yields of the RS and WS positrons from each of the ST mode are very limited in each of the momentum bins. 
From separate fits to $M_{\rm BC}$ distributions in each data set, the yields of the total RS and WS positrons yields are determined to be $3706\pm71$ and $394\pm31$, respectively, where the uncertainties are statistical only. 
The measured yields as a function of momentum bin for each particle category are shown in Fig.~\ref{fig:Nobs}.

\subsection{PID unfolding}
The measured positron yields in each momentum bin contain sizable backgrounds from misidentified hadrons. To evaluate these backgrounds, knowledge of their yields and corresponding identification and misidentification probabilities is required.
We relate the true number of positrons and the observed number of positrons in DT events with a PID migration matrix ($A_{\rm PID}$):
\begin{equation}
\left[
\begin{array} {ccc}
N^{\rm obs}_{e}  \\
N^{\rm obs}_{\pi}  \\
N^{\rm obs}_{K} \\
N^{\rm obs}_{p} \\
\end{array}
\right]
=
\left[
\begin{array} {cccc}
P_{e\rightarrow e}    & P_{\pi\rightarrow e}    & P_{K\rightarrow e}    & P_{p\rightarrow e}  \\
P_{e\rightarrow \pi}  & P_{\pi\rightarrow \pi}  & P_{K\rightarrow \pi}  & P_{p\rightarrow \pi} \\
P_{e\rightarrow K}   & P_{\pi\rightarrow K}    & P_{K\rightarrow K}   & P_{p\rightarrow K}  \\
P_{e\rightarrow p}   & P_{\pi\rightarrow p}    & P_{K\rightarrow p}    & P_{p\rightarrow p}  \\
\end{array}
\right]
\left[
\begin{array} {ccc}
N^{\rm true}_{e}  \\
N^{\rm true}_{\pi}  \\
N^{\rm true}_{K} \\
N^{\rm true}_{p} \\
\end{array}
\right],
\end{equation}
where $N^{\rm obs}_{a}$ is the observed yield of particle species $a$ (where $a$ denotes $e$, $\pi$, $K$ or $p$), $P_{a\rightarrow b}$ is the efficiency of identifying particle $a$ as particle $b$, and $N^{\rm true}_{a}$ is true yield of particle $a$ in the studied sample.  We do not attempt to separate muons; most muons will be identified as pions by our classification scheme.  
To obtain the elements in the PID efficiency matrix ($A_{\rm PID}$), the control samples for positrons, charged pions, kaons and protons are studied.
These control samples are selected from the $\Lambda_c^+$ inclusive MC sample, by comparing the opening angle ($\Delta_{\rm angle}<5^{\circ}$) between the reconstructed charged tracks and the produced charged particles. 
Based on these selected control samples of positrons, pions, kaons and protons, the elements of $A_{\rm PID}$, $P_{a\rightarrow b}$, as a function of momentum are obtained. 
To account for the possible differences in the migration probabilities $P_{a\rightarrow b}$ between data and MC sample, corrections to $P_{a\rightarrow b}$ are applied in each momentum bin. 
For positrons, the difference between the data and the MC sample is studied by requiring the positron PID efficiency obtained from radiative Bhabha scattering process. For kaons and pions, the differences in PID efficiencies between data and MC sample are studied using the decays 
$J/\psi\rightarrow K^+K^-\pi^+\pi^-(\pi^0)$ and $J/\psi\rightarrow K^+K^-K^+K^-(\pi^0)$. For protons, the decay $J/\psi\rightarrow p\bar{p}\pi^+\pi^-(\pi^0)$ is analyzed.
The averaged relative differences (uncertainties) are evaluated to be $-0.5(0.4)\%$, $-2.2(0.3)\%$, $-1.0(0.2)\%$ and $-2.3(0.3)\%$ for $P_{e\rightarrow e}$, $P_{\pi\rightarrow \pi}$, $P_{K\rightarrow K}$ and $P_{p\rightarrow p}$, respectively.
After these corrections, the PID efficiencies for each of the charged particles as a function of momentum are shown in Fig.~\ref{fig:APID}. An example of the $A_{\rm PID}$ for $0.50<p<0.55$~GeV/$c$ is shown in the Appendix A.

\begin{figure}[tp!]
\begin{center}
   \begin{minipage}[t]{8cm}
   \includegraphics[width=\linewidth]{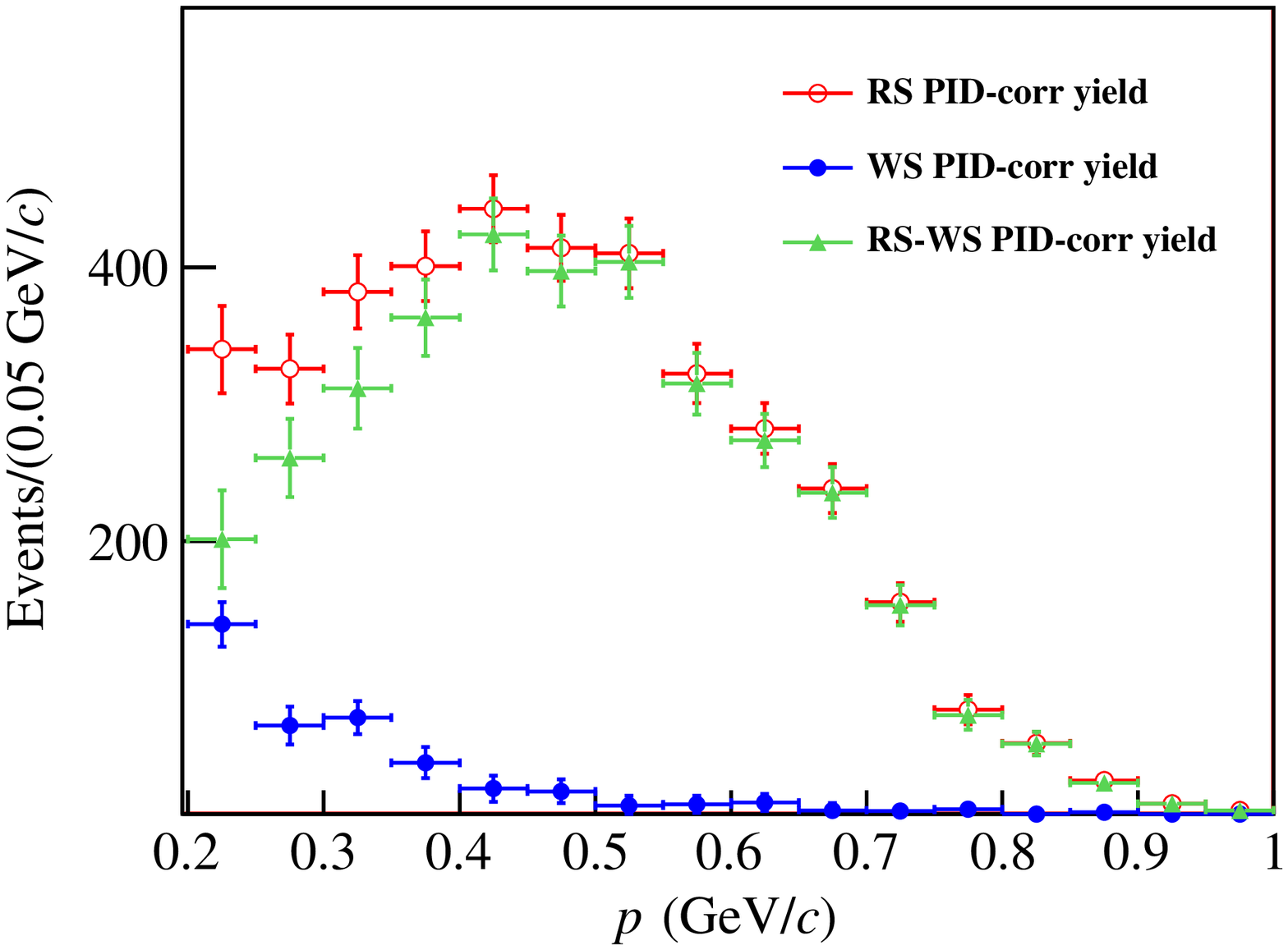}
   \end{minipage}
   \caption{PID unfolding results of positrons from $\Lambda_c^+\rightarrow Xe^+\nu_e$ events for data sets collected between $\sqrt{s}$=4.600~GeV and 4.698~GeV. The RS PID-corrected yields are shown in red circles, while the WS PID-corrected yields are shown in blue dots. Their differences are shown as green triangles. } 
\label{fig:PIDun}
\end{center}
\end{figure}

We unfold PID migration effects by applying the inverse of the migrations matrix to the observed RS and WS yields in each momentum bin.  
We obtain the true yield of RS positron tracks from $\Lambda_c^+$ decays and the true WS backgrounds, as shown in Fig.~\ref{fig:PIDun}. 
We randomly Gaussian smear each observed yield with its uncertainty and repeat the unfolding procedure many times. The RMS of each resulting unfolded yield is assigned as its statistical uncertainty.   
We then take the difference of the number of the RS positrons and WS electrons to subtract the contributions of positrons from Dalitz decays of light mesons in the final state of $\Lambda_c^+$ decay, such as $\pi^0\rightarrow \gamma e^+e^-$, and any other charge-symmetric backgrounds.

\begin{figure*}[tp!]
\begin{center}
   \begin{minipage}[t]{18cm}
   \includegraphics[width=\linewidth]{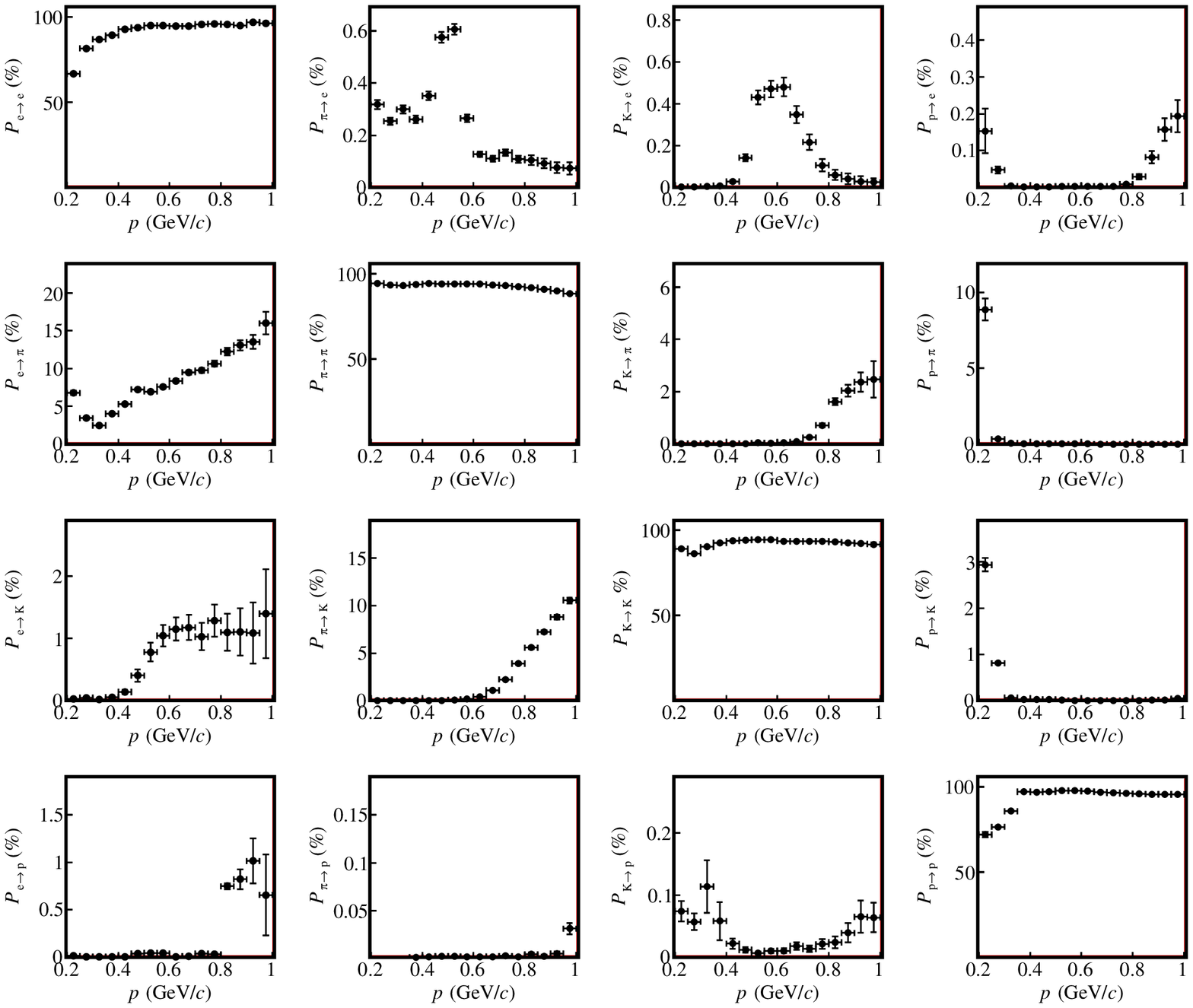}
   \end{minipage}
   \caption{ PID efficiencies as a function of momentum used to populate the $A_{\rm PID}$ matrices.}
\label{fig:APID} 
\end{center}
\end{figure*}

%

\subsection{Tracking Efficiency and Momentum Migration}
To account for tracking efficiency and momentum bin migration, a second unfolding is performed using 
\begin{equation}
N_{\rm e}^{\rm true}(i)=\sum_j A_{\rm TRK}(e|i,j)N_{\rm e}^{\rm prod}(j),
\label{eq:formula}
\end{equation}
where the tracking efficiency matrix $A_{\rm TRK}$ describes the efficiency that positrons produced in the $j$th momentum bin are reconstructed in the $i$th momentum bin, $N_{\rm e}^{\rm prod}(j)$ is the number of primary positrons produced in the $j$th momentum bin, and $N_{\rm e}^{\rm true}(i)$ is the true yield of positron reconstructed in the $i$th momentum bin.


To determine the tracking efficiency matrix $A_{\rm TRK}$, the MC samples of one $\Lambda_c^+$ baryon decaying to $Xe^+\nu_e$ together with a $\bar{\Lambda}_c$ decaying to each ST mode are generated at each of the center-of-mass energies. The tracking efficiency of $\Lambda_c^+\rightarrow Xe^+\nu_e$ for each ST mode is determined firstly and then weighted according to their ST yields. The tracking efficiency matrix $A_{\rm TRK}$ is shown in the Appendix A, including the effects of geometrical acceptance ($|\!\cos\theta|<0.80$), track reconstruction efficiency, and resolution smearing. Unfolding with the matrix inverse, we obtain the efficiency-corrected positron momentum spectrum above 200~MeV/$c$ in the laboratory system. The corrected positron yield is determined to be $4333\pm107$, where the uncertainty is statistical only. The efficiency-corrected positron yields in each of the momentum bins as well as their correlation coefficients are shown in the Appendix A.
Table~\ref{tab:Nobs} summarizes the positron yields obtained after each correction step. 

\begin{table}
\caption{ Positron yield in data after each procedure. The listed uncertainties are statistical. }
\begin{center}
\begin{tabular}
{lcc} \hline\hline
Correction (see text)     &  RS yields    & WS yields \\
\hline
Observed yields           &  $3706\pm 71$ & $394\pm31$ \\
PID unfolding yields      &  $3865\pm 80$ & $376\pm33$   \\
WS subtraction            &  $3489\pm 87$ \\   
Tracking unfolding yields &  $4333\pm107$ \\
Extrapolation             &  $4692\pm117$ \\
\hline\hline
\end{tabular}
\label{tab:Nobs}
\end{center}
\end{table}

\begin{figure}[tp!]
\begin{center}
   \begin{minipage}[t]{8cm}
   \includegraphics[width=\linewidth]{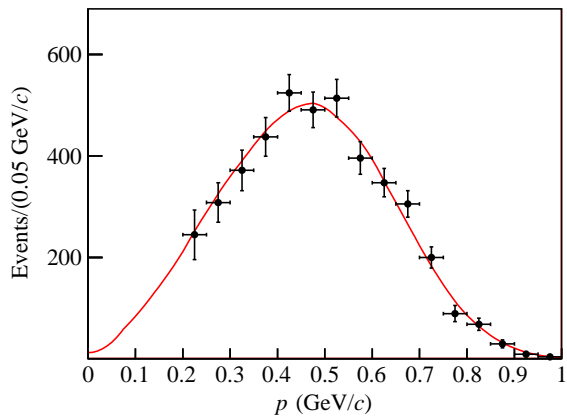}
   \end{minipage}
   \caption{ Extrapolation of the positron momentum spectrum  obtained from the data. The black points show data, while the red curve shows the extrapolated momentum spectrum.}
\label{fig:TRKun}
\end{center}
\end{figure}

\begin{table}
\caption{Exclusive $\Lambda_c^+$ SL decays used to extrapolate the positron momentum spectrum. The BFs of $\Lambda_c^+\rightarrow \Lambda e^+\nu_e$ and $\Lambda_c^+\rightarrow pK^-e^+\nu_e$ are measured by BESIII experiment~\cite{Lamev,pKev}. The BF of $\Lambda_c^+\rightarrow n\bar{K}^0e^+\nu_e$ is taken as the BF of $\Lambda_c^+\rightarrow pK^-e^+\nu_e$ according to isospin symmetry of $N\bar{K}$ system. 
The form factor of the $\Lambda_c^+\rightarrow \Lambda e^+\nu_e$ is taken from BESIII measurement~\cite{Lamev}. The model for $\Lambda_c^+\rightarrow pK^-(n\bar{K}^0)e^+\nu_e$ is taken as phase space (PHSP).
The BFs and models of $\Lambda_c^+\rightarrow \Lambda(1405)e^+\nu_e$ and $\Lambda_c^+\rightarrow \Lambda(1520)e^+\nu_e$ are taken from the predictions in Ref.~\cite{PRD95_053005,PRC72_035201}, while the BF and model of $\Lambda_c^+\rightarrow ne^+\nu_e$ are taken from Ref.~\cite{PRD90_114033}.  }
\begin{center}
\begin{tabular}
{lccc} \hline\hline
Decay  & $\mathcal{B}$ [\%] & Model  \\
\hline $\Lambda_c^+\rightarrow \Lambda e^+\nu_e$ & $3.56\pm0.11\pm0.07$ & Ref.~\cite{Lamev} \\
\hline $\Lambda_c^+\rightarrow pK^-(n\bar{K}^0)e^+\nu_e$ & $0.088\pm0.017\pm0.007$ & PHSP~\cite{pKev} \\
\hline $\Lambda_c^+\rightarrow \Lambda(1405)e^+\nu_e$  & 0.24 & HQET~\cite{PRD95_053005,PRC72_035201}  \\
\hline $\Lambda_c^+\rightarrow \Lambda(1520)e^+\nu_e$  & $0.06$ & HQET~\cite{PRD95_053005,PRC72_035201}   \\
\hline $\Lambda_c^+\rightarrow ne^+\nu_e$ & $0.20$ & Quark model~\cite{PRD90_114033} \\
\hline\hline
\end{tabular}
\label{tab:Model}
\end{center}
\end{table}

\subsection{Branching fraction}
The yield of positrons with $p_e\le200$~MeV/$c$ is obtained by fitting the efficiency-corrected positron momentum spectrum with the sum of the spectra of the exclusive decay channels, 
as shown in Fig.~\ref{fig:TRKun}. In the fit, the BF of each component is fixed to its central value. The details of the incorporated exclusive channels are shown in Table~\ref{tab:Model}. 
Combining with the measured yields with $p_e>200$~MeV/$c$, we obtain the total efficiency-corrected yield of $\Lambda_c^+\rightarrow Xe^+\nu_e$, $N_{\rm prod}=4692\pm117$,  where the uncertainty is statistical only.
This allows us to calculate the BF of $\Lambda_c^+\rightarrow Xe^+\nu_e$ by:
\begin{equation}
\mathcal{B}(\Lambda_c^+\rightarrow Xe^+\nu_e)=\frac{N_{\rm prod}}{N_{\rm ST}}.
\label{eq:Bran}
\end{equation}
Inserting $N_{\rm ST}=115437\pm446$ into Eq.~(\ref{eq:Bran}), we measure 
$$\mathcal{B}(\Lambda_c^+\rightarrow Xe^+\nu_e)=(4.06\pm0.10_{\rm stat.}\pm0.09_{\rm syst})\%.$$
The measured $\mathcal{B}(\Lambda_c^+\rightarrow Xe^+\nu_e)$ is consistent with the previous result $\mathcal{B}(\Lambda_c^+\rightarrow Xe^+\nu_e)=(3.95\pm0.34_{\rm stat.}\pm0.09_{\rm syst})\%$~\cite{PRL121_251801}, but with 
greatly improved precision.  

\subsection{Systematic uncertainty}

The systematic uncertainties in measuring $\mathcal{B}(\Lambda_c^+\rightarrow Xe^+\nu_e)$ are listed in Table~\ref{tab:SystErr}. The uncertainty in the tracking efficiency of positron is evaluated at 0.4\% from studies of $e^+e^-\rightarrow (\gamma)e^+e^-$ events. The systematic uncertainty in the ST yields is estimated to be 1.0\% studied using alternative signal and background shapes~\cite{Lamev}.
The systematic uncertainty due to the $A_{\rm PID}$ and $A_{\rm TRK}$ matrices is estimated by randomly Gaussian smearing all elements of these matrices 300 times, based on their errors, and re-determining the BF. The 0.9\% width of the resulting distribution of these BF results is taken as the systematic uncertainty. 
The systematic uncertainty in the extrapolation of positron momentum spectrum is estimated by varying the BFs of $\Lambda_c^+\rightarrow \Lambda e^+\nu_e$, $\Lambda_c^+\rightarrow pK^-e^+\nu_e$, $\Lambda_c^+\rightarrow n\bar{K}^0e^+\nu_e$, $\Lambda_c^+\rightarrow \Lambda(1405)e^+\nu_e$, $\Lambda_c^+\rightarrow \Lambda(1520)e^+\nu_e$ and $\Lambda_c^+\rightarrow ne^+\nu_e$. The BFs of $\Lambda_c^+\rightarrow \Lambda e^+\nu_e$ and $\Lambda_c^+\rightarrow pK^-(n\bar{K}^0)e^+\nu_e$ are varied by one standard deviation according to measurements~\cite{Lamev,pKev}. The BFs for $\Lambda_c^+\rightarrow \Lambda(1405)e^+\nu_e$ and $\Lambda_c^+\rightarrow \Lambda(1520)e^+\nu_e$ are varied from 0.24\% to 0.38\%, and from 0.06\% to 0.08\% according to predictions in Refs.~\cite{PRD95_053005,PRC72_035201}. The BF for $\Lambda_c^+\rightarrow ne^+\nu_e$ is varied from 0.2\% to 0.4\% according to predictions in Refs.~\cite{PRC72_035201,PRD97_034511,PRD56_348,PRD90_114033,PRD93_056008,EPJC76_628,JPG44_075006}. In addition, there may be multi-body decays like  $\Lambda_c^+\rightarrow \Lambda\pi^0e^+\nu_e$, $\Lambda_c^+\rightarrow \Lambda\pi^+\pi^-e^+\nu_e$, etc. In the absence of data or predictions, the BFs of these decays 
are both taken as the conservatively large value 0.2\%. 
With these exclusive BFs varied one by one in alternative fits, the signal shape is reformed and the $\mathcal{B}(\Lambda_c^+\rightarrow Xe^+\nu_e)$ is obtained. 
The largest relative difference, 1.6\%, between the new measured BF 
and the nominal one is taken as the systematic uncertainty
due to momentum extrapolation.  
In addition, the observed yields of the RS and WS positrons in different momentum bins are determined by fitting to $M_{\rm BC}$ distributions obtained by combining twelve hadronic ST modes together. 
There may be a systematic difference if instead each ST mode is fit separately 
and then the yields summed.  This difference is investigated by studying data collected at $\sqrt{s}=4.682$~GeV. We compared the total ST yields with the two methods and the relative difference of 0.6\% is taken as systematic uncertainty.  Next, in the PID unfolding procedure, muons are treated as pions due to their similar detector responses, which may introduce a bias. To consider this systematic uncertainty, the yield of the $\Lambda_c^+\rightarrow X\mu^+\nu_{\mu}$ component is estimated and the fake rates of muons to charged positrons, pions, kaons and protons are studied by using MC samples. Then we redo the analysis procedure and take the relative difference of 0.2\% as the systematic uncertainty due to the treatment of muons. Summing in quadrature, we determine the total systematic uncertainty in measuring $\mathcal{B}(\Lambda_c^+\rightarrow Xe^+\nu_e)$ to be 2.2\%.

\begin{table}
\caption{ Sources of the systematic uncertainties in measuring $\mathcal{B}(\Lambda_c^+\rightarrow Xe^+\nu_e)$. }
\begin{center}
\begin{tabular}
{l|c} \hline\hline
Source ~~~~~~~~~~~~~~~~~~~~~~~~~~~~~~~~~~~~~~~~~~~~~& Value  \\
\hline Tracking efficiency for positron & 0.4\% \\
\hline ST signal shape  & 1.0\% \\
\hline $A_{\rm PID}$ and $A_{\rm TRK}$ matrices & 0.9\% \\
\hline Momentum extrapolation & 1.6\% \\
\hline ST Yields method for RS and WS positrons & 0.6\% \\
\hline Muon contamination treatment & 0.2\% \\
\hline Total & 2.2\% \\
\hline\hline
\end{tabular}
\label{tab:SystErr}
\end{center}
\end{table}

\section{Summary}
Based on analyzing 4.5 fb$^{-1}$ data taken at $\sqrt{s}=4.600$, $4.612$, $4.628$, $4.640$, $4.661$, $4.682$ and $4.698$~GeV with the BESIII detector at the BEPCII collider, the inclusive SL decay 
$\Lambda_c^+\rightarrow Xe^+\nu_e$ is investigated. The BF of the decay is measured to be $\mathcal{B}({\it \Lambda}^+_c\rightarrow Xe^+\nu_e)=(4.06\pm0.10_{\rm stat.}\pm0.09_{\rm syst.})\%$. 
Combining the lifetime of $\Lambda_c^+$ baryon $\tau_{\Lambda_c^+}=(202.4\pm3.1)\times 10^{-15}$~s~\cite{pdg2020}, we obtain the decay width $\Gamma(\Lambda_c^+\rightarrow X e^+\nu_e)=(2.006\pm0.073)\times 10^{11}$~s$^{-1}$. The charge-averaged value for $D$ inclusive SL decay width is determined to be $\bar{\Gamma}(D\rightarrow Xe^+\nu_e)=(1.567\pm0.020)\times 10^{11}$~s$^{-1}$~\cite{pdg2020}. 
Hence, the ratio of the inclusive SL decay width for the $\Lambda_c^+$ and $D$ is:
$$\frac{\Gamma(\Lambda_c^+\rightarrow X e^+\nu_e)}{\bar{\Gamma}(D\rightarrow Xe^+\nu_e)}=1.28\pm0.05,$$
which is consistent but greatly improved compared with the previous published result $\Gamma(\Lambda_c^+\rightarrow \Lambda e^+\nu_e)/\bar{\Gamma}(D\rightarrow Xe^+\nu_e)=1.26\pm0.12$~\cite{PRL121_251801}.  At a confidence level of 95\%, 
the determined ratio, $\Gamma(\Lambda_c^+\rightarrow \Lambda e^+\nu_e)/\bar{\Gamma}(D\rightarrow Xe^+\nu_e)=1.28\pm0.05$, is consistent with the value of 1.2 predicted from the heavy quark expansion~\cite{PRD49_1310} but disfavors the value of 1.67 predicted from the effective-quark method~\cite{PRD83_034025,PRD86_014017}.

The BESIII collaboration thanks the staff of BEPCII and the IHEP computing center for their strong support. This work is supported in part by National Key R\&D Program of China under Contracts Nos. 2020YFA0406400, 2020YFA0406300; National Natural Science Foundation of China (NSFC) under Contracts Nos. 11635010, 11735014, 11835012, 11935015, 11935016, 11935018, 11961141012, 12022510, 12025502, 12035009, 12035013, 12192260, 12192261, 12192262, 12192263, 12192264, 12192265; the Chinese Academy of Sciences (CAS) Large-Scale Scientific Facility Program; Joint Large-Scale Scientific Facility Funds of the NSFC and CAS under Contract No. U1832207; 100 Talents Program of CAS; The Institute of Nuclear and Particle Physics (INPAC) and Shanghai Key Laboratory for Particle Physics and Cosmology; ERC under Contract No. 758462; European Union's Horizon 2020 research and innovation programme under Marie Sklodowska-Curie grant agreement under Contract No. 894790; German Research Foundation DFG under Contracts Nos. 443159800, Collaborative Research Center CRC 1044, GRK 2149; Istituto Nazionale di Fisica Nucleare, Italy; Ministry of Development of Turkey under Contract No. DPT2006K-120470; National Science and Technology fund; National Science Research and Innovation Fund (NSRF) via the Program Management Unit for Human Resources \& Institutional Development, Research and Innovation under Contract No. B16F640076; STFC (United Kingdom); Suranaree University of Technology (SUT), Thailand Science Research and Innovation (TSRI), and National Science Research and Innovation Fund (NSRF) under Contract No. 160355; The Royal Society, UK under Contracts Nos. DH140054, DH160214; The Swedish Research Council; U. S. Department of Energy under Contract No. DE-FG02-05ER41374

\newpage
\appendix
\section{Tables for $A_{\rm TRK}$ and efficiency-corrected yields}
\label{sec:appendixa}

Table~\ref{tab:APID_0525} shows an example of the PID efficiency matrix $A_{\rm PID}$ for momentum region within $0.50<p<0.55$~GeV/$c$. Table~\ref{tab:trackeff} shows the tracking efficiency matrix $A_{\rm TRK}$ of positron for each of the momentum bins. Table.~\ref{tab:yield} shows the efficiency-corrected positron yields in each of the momentum bins and their correlation coefficients.

\begin{table}
\caption{ The PID efficiency matrix $A_{\rm PID}$ (in \%) for $0.50<p<0.55$~GeV/$c$. }
\begin{center}
\begin{tabular}
{c|rrrr} \hline\hline
$P_{a\rightarrow b}$ (\%) & $a=e$ & $a=\pi$ & $a=K$ & $a=p$ \\
\hline 
$b=e$   & 95.17 &   0.61 &   0.43 &   0.00  \\
$b=\pi$ &   6.96 & 93.80 &   0.04 &   0.01  \\
$b=K$  &   0.78 &   0.07 & 94.42 &   0.01  \\
$b=p$  &   0.04 &   0.00 &   0.01 & 97.80 \\
\hline\hline
\end{tabular}
\label{tab:APID_0525}
\end{center}
\end{table}

\normalsize
 
\begin{table*}
\caption{ \small The tracking efficiency matrix $A_{\rm TRK}$ for positron. The column gives the true momentum bins $j$, while
the row gives the reconstructed bin $i$.} \scriptsize
\begin{center}
\begin{tabular}
{|c|cccccccccccccccc|} \hline\hline
$A_{\rm TRK}(e|i,j)$  & 1  & 2  & 3  & 4  & 5  & 6  & 7  & 8  & 9  & 10 & 11 & 12 & 13 & 14 & 15 & 16  \\
\hline
 1 & 0.7223&0.0338&0.0089&0.0058&0.0037&0.0029&0.0028&0.0024&0.0024&0.0022&0.0022&0.0022&0.0030&0.0032&0.0036&0.0009 \\
 2 & 0.0174&0.7359&0.0383&0.0097&0.0055&0.0041&0.0034&0.0033&0.0031&0.0027&0.0027&0.0034&0.0029&0.0034&0.0067&0.0105 \\
 3 & 0.0008&0.0139&0.7372&0.0389&0.0106&0.0063&0.0047&0.0041&0.0035&0.0031&0.0031&0.0032&0.0029&0.0031&0.0032&0.0017 \\
 4 & 0.0004&0.0008&0.0154&0.7355&0.0396&0.0106&0.0067&0.0051&0.0041&0.0036&0.0039&0.0029&0.0033&0.0038&0.0039&0.0012 \\
 5 & 0.0002&0.0003&0.0006&0.0148&0.7311&0.0412&0.0116&0.0070&0.0054&0.0047&0.0045&0.0035&0.0037&0.0037&0.0033&0.0014 \\
 6 & 0.0001&0.0003&0.0002&0.0007&0.0155&0.7248&0.0432&0.0118&0.0071&0.0056&0.0045&0.0038&0.0038&0.0050&0.0034&0.0043 \\
 7 & 0.0002&0.0002&0.0002&0.0003&0.0005&0.0173&0.7175&0.0455&0.0118&0.0077&0.0054&0.0049&0.0042&0.0045&0.0039&0.0028 \\
 8 & 0.0001&0.0001&0.0001&0.0002&0.0003&0.0008&0.0180&0.7125&0.0481&0.0120&0.0080&0.0056&0.0050&0.0044&0.0062&0.0014 \\
 9 & 0.0001&0.0001&0.0001&0.0002&0.0002&0.0004&0.0008&0.0196&0.7069&0.0507&0.0126&0.0084&0.0054&0.0048&0.0037&0.0018 \\
10 & 0.0001&0.0001&0.0000&0.0001&0.0001&0.0002&0.0004&0.0006&0.0200&0.7043&0.0536&0.0119&0.0083&0.0047&0.0045&0.0020 \\
11 & 0.0000&0.0000&0.0000&0.0001&0.0001&0.0001&0.0002&0.0003&0.0006&0.0208&0.6982&0.0580&0.0129&0.0099&0.0093&0.0047 \\
12 & 0.0000&0.0000&0.0000&0.0001&0.0001&0.0001&0.0002&0.0002&0.0005&0.0004&0.0223&0.7041&0.0631&0.0117&0.0107&0.0064 \\
13 & 0.0000&0.0000&0.0000&0.0000&0.0001&0.0000&0.0001&0.0001&0.0002&0.0003&0.0006&0.0213&0.6950&0.0682&0.0146&0.0046 \\
14 & 0.0000&0.0000&0.0000&0.0000&0.0001&0.0000&0.0000&0.0001&0.0001&0.0002&0.0002&0.0003&0.0218&0.6997&0.0823&0.0077 \\
15 & 0.0000&0.0000&0.0000&0.0000&0.0000&0.0000&0.0000&0.0000&0.0001&0.0001&0.0002&0.0003&0.0003&0.0228&0.7165&0.0712 \\
16 & 0.0000&0.0000&0.0000&0.0000&0.0000&0.0000&0.0000&0.0000&0.0000&0.0001&0.0001&0.0001&0.0001&0.0005&0.0223&0.6855 \\
\hline\hline
\end{tabular}
\label{tab:trackeff}
\end{center}
\end{table*}

\begin{table*}
\caption{ The efficiency-corrected positron yields $N^{\rm prod}_e(i)$ in each of momentum bins and their correlation coefficients.}
\begin{center}
\begin{tabular}{c|c|rrrrrrrrrrrrrrrr}
\hline\hline
bins~ & ~~$N^{\rm prod}_e(i)$~~ & 1~~ & 2~~ & 3~~ & 4~~ & 5~~ & 6~~ & 7~~ & 8~~ & 9~~ & 10~ & 11~ & 12~ & 13~ & 14~ & 15~ \\
    \hline
 1 & $244\pm49$ &    \\
 2 & $308\pm39$ & -0.07 &    \\
 3 & $371\pm40$ & -0.00 & -0.07 &    \\
 4 & $437\pm38$ & -0.01 & -0.00 & -0.07 &    \\
 5 & $524\pm36$ & -0.02 & -0.01 & -0.00 & -0.07 &    \\
 6 & $491\pm35$ &  0.00 & -0.02 & -0.01 & -0.00 & -0.08 &    \\
 7 & $513\pm37$ & -0.00 &  0.00 & -0.02 & -0.01 & -0.00 & -0.09 &    \\
 8 & $395\pm32$ & -0.00 & -0.00 &  0.00 & -0.02 & -0.01 & -0.00 & -0.09 &   \\
 9 & $347\pm28$ & -0.00 & -0.00 & -0.00 &  0.00 & -0.02 & -0.01 & -0.00 & -0.09 &    \\
10 & $305\pm26$ &  0.01 & -0.00 & -0.00 & -0.00 &  0.00 & -0.02 & -0.01 & -0.00 & -0.10 &    \\
11 & $199\pm21$ & -0.01 &  0.01 & -0.00 & -0.00 & -0.00 &  0.00 & -0.02 & -0.01 &  0.00 & -0.10 &    \\
12 & $ 89\pm16$ &  0.01 & -0.01 &  0.01 & -0.00 & -0.00 & -0.00 &  0.00 & -0.02 & -0.01 &  0.00 & -0.11 &   \\
13 & $ 68\pm12$ & -0.01 &  0.01 & -0.01 &  0.01 & -0.00 & -0.00 & -0.00 &  0.00 & -0.02 & -0.01 &  0.00 & -0.11 &     \\
14 & $ 29\pm 8~$ &  0.00 & -0.01 &  0.01 & -0.01 &  0.01 & -0.00 & -0.00 & -0.00 &  0.00 & -0.01 & -0.01 &  0.01 & -0.11 &  &    \\
15 & $  9\pm 5$ & -0.01 &  0.00 & -0.01 &  0.01 & -0.01 &  0.01 & -0.00 & -0.00 & -0.00 &  0.00 & -0.02 & -0.01 &  0.01 & -0.13 &    \\
16 & $  4\pm 3$ &  0.00 & -0.01 &  0.00 & -0.01 &  0.01 & -0.01 &  0.01 &  0.00 & -0.00 & -0.00 &  0.00 & -0.01 & -0.01 &  0.01 & -0.10   \\
\hline\hline
\end{tabular}
\label{tab:yield}
\end{center}
\end{table*}


\begin{thebibliography}{99}

\bibitem{PRL37_882} B. Knapp $et~al.$, \href{https://doi.org/10.1103/PhysRevLett.37.882}{Phys. Rev. Lett. {\bf 37}, 882 (1976)}.
\bibitem{PRL44_10} G. S. Abrams $et~al.$ [Mark II Collaboration], \href{https://doi.org/10.1103/PhysRevLett.44.10}{Phys. Rev. Lett. 44, 10 (1980)}.
\bibitem{pdg2020} P.A. Zyla et al. (Particle Data Group), \href{https://pdg.lbl.gov/}{Prog. Theor. Exp. Phys. 2020, 083C01 (2020) and 2021 online update}. 
\bibitem{CPC44_040001} M.~Ablikim $et~al.$ [BESIII Collaboration], \href{https://doi.org/10.1088/1674-1137/44/4/040001}{Chin. Phys. C {\bf 44}, 040001 (2020)}.
\bibitem{NSR8_11} H.~B.~Li and X.~R.~Lyu, \href{https://doi.org/10.1093/nsr/nwab181}{Natl. Sci. Rev. \textbf{8}, no.~11, nwab181 (2021)}. 

\bibitem{Lamev} M.~Ablikim $et~al.$ [BESIII Collaboration], \href{https://doi.org/10.1103/PhysRevLett.129.231803}{Phys. Rev. Lett. 129, 231803 (2022)}.
\bibitem{pKev} M.~Ablikim $et~al.$ [BESIII Collaboration],  \href{https://doi.org/10.48550/arXiv.2207.11483}{arXiv:2207.11483}. 
\bibitem{PRL121_251801} M.~Ablikim $et~al.$ [BESIII Collaboration],  \href{https://doi.org/10.1103/PhysRevLett.121.251801}{Phys. Rev. Lett. {\bf 121}, 251801 (2018)}. 

\bibitem{PRD83_034025} M. Gronau and J. L. Rosner, \href{https://doi.org/10.1103/PhysRevD.83.034025}{Phys. Rev. D {\bf 83}, 034025 (2011)}.
\bibitem{PRD86_014017} J. L. Rosner, \href{https://doi.org/10.1103/PhysRevD.86.014017}{Phys. Rev. D {\bf 86}, 014017 (2012)}.
\bibitem{PRD49_1310} A. V. Manohar and M. B. Wise, \href{https://doi.org/10.1103/PhysRevD.49.1310}{Phys. Rev. D {\bf 49}, 1310 (1994)}.


\bibitem{lum_4600} M.~Ablikim $et~al.$ [BESIII Collaboration], \href{https://doi.org/10.1088/1674-1137/39/9/093001}{Chin. Phys. C {\bf 39}, 093001 (2015)}.
\bibitem{lum_new} M.~Ablikim $et~al.$ [BESIII Collaboration], \href{https://doi.org/10.1088/1674-1137/ac84cc}{Chin. Phys. C {\bf 46}, 113003 (2022)}.


\bibitem{Ablikim:2009aa} M.~Ablikim $et~al.$ [BESIII Collaboration], \href{https://doi.org/10.1016/j.nima.2009.12.050}{Nucl.\ Instrum.\ Meth.\ A {\bf 614}, 345 (2010)}.

\bibitem{Muon} K.~X.~Huang, $et~al.$, \href{https://doi.org/10.1007/s41365-022-01133-8}{Nucl. Sci. Tech. {\bf}33, 142 (2022).} 

\bibitem{tofup} X. Li $et~al.$, Radiat. Detect. Technol. Methods {\bf 1}, 13 (2017); Y. X. Guo $et~al.$, Radiat. Detect.
Technol. Methods {\bf 1}, 15 (2017); P. Cao $et~al.$, \href{https://doi.org/10.1016/j.nima.2019.163053}{Nucl. Instrum. Meth. A {\bf 953}, 163053 (2020)}.

\bibitem{geant4} S.~Agostinelli $et~al.$ [GEANT4 Collaboration], \href{http://dx.doi.org/10.1016/S0168-9002(03)01368-8}{Nucl. Instrum. Meth. A {\bf 506}, 250 (2003)}.
\bibitem{SJNP41_466} E.~A.~Kurav and V.~S.~Fadin, Sov. J. Nucl. Phys. {\bf 41}, 466 (1985).
\bibitem{kkmc} S. Jadach, B. F. L. Ward and Z. Was, \href{https://doi.org/10.1016/S0010-4655(00)00048-5}{Comput. Phys. Commun. {\bf 130}, 260 (2000)}; \href{https://doi.org/10.1103/PhysRevD.63.113009}{Phys. Rev. D {\bf 63}, 113009 (2001)}.
\bibitem{plb303_163} E.~Richter-Was, \href{https://doi.org/10.1016/0370-2693(93)90062-M}{Phys. Lett. B {\bf 303}, 163 (1993)}.
\bibitem{nima462_152} D.~J.~Lange, \href{https://doi.org/10.1016/S0168-9002(01)00089-4}{Nucl. Instrum. Meth. A {\bf 462}, 152 (2001)}; \href{https://doi.org/10.1088/1674-1137/32/8/001}{R. G. Ping, Chin. Phys. C {\bf 32}, 599 (2008)}.


\bibitem{lundcharm} J.~C.~Chen, G.~S.~Huang, X.~R.~Qi, D.~H.~Zhang and Y.~S.~Zhu, \href{https://doi.org/10.1103/PhysRevD.62.034003}{Phys. Rev. D {\bf 62}, 034003 (2000)}.
\bibitem{lundcharm2} R.~L.~Yang, R.~G.~Ping and H.~Chen, \href{http://cpl.iphy.ac.cn/EN/article/searchArticleResult.do}{Chin. Phys. Lett. {\bf 31} 061301 (2014)}.

\bibitem{ST} J. Adler $et~al.$ [Mark III Collaboration],  \href{https://doi.org/10.1103/PhysRevLett.121.251801}{Phys. Rev. Lett. {\bf 62}, 1821 (1989)}. 
\bibitem{PRD104_012003} M.~Ablikim $et~al.$ [BESIII Collaboration], \href{https://doi.org/10.1103/PhysRevD.104.012003}{Phys. Rev. D {\bf 104}, 012003 (2021)}.

\bibitem{plb241_278} H. Albrecht $et~al.$ [ARGUS Collaboration], \href{https://doi.org/10.1016/0370-2693(90)91293-K}{Phys. Lett. B. {\bf 241}, 278 (1990)}.

\bibitem{PRD95_053005} M.~M.~Hussain and W.~Roberts, \href{https://doi.org/10.1103/PhysRevD.95.053005}{Phys. Rev. D {\bf 95}, 053005 (2017)}; \href{https://doi.org/10.1103/PhysRevD.95.099901}{Phys. Rev. D {\bf 95}, 099901 (2017)}.
\bibitem{PRC72_035201} M. Pervin, W. Roberts and S. Capstick, \href{https://doi.org/10.1103/PhysRevC.72.035201}{Phys. Rev. C {\bf 72}, 035201 (2005)}. 


\bibitem{PRD90_114033} T. Gutsche, M. A. Ivanov, J. G. K{\"o}rner, V. E. Lyubovitskij, and P. Santorelli, \href{https://doi.org/10.1103/PhysRevD.90.114033}{Phys. Rev. D {\bf 90}, 114033 (2014)}; Erratum, \href{https://doi.org/10.1103/PhysRevD.94.059902}{Phys. Rev. D {\bf 94}, 059902 (2016)}.
\bibitem{PRD97_034511} Stefan Meinel, \href{https://doi.org/10.1103/PhysRevD.97.034511}{Phys. Rev. D {\bf 97}, 034511 (2018)}.

\bibitem{PRD56_348} M. A. Ivanov, V. E. Lyubovitskij, J. G. K{\"o}rner, and P. Kroll, \href{https://doi.org/10.1103/PhysRevD.56.348}{Phys. Rev. D {\bf 56}, 348 (1997)}.

\bibitem{PRD93_056008} C.-D. L{\"u}, W. Wang, and F.-S. Yu, \href{10.1103/PhysRevD.93.056008}{Phys. Rev. D {\bf 93}, 056008 (2016)}.
\bibitem{EPJC76_628} R.~N. Faustov, V.~O.~Galkin, \href{https://doi.org/10.1140/epjc/s10052-016-4492-z}{Eur. Phys. J. C {\bf 76} 628 (2016)}.
\bibitem{JPG44_075006} C.~F. Li, Y.~L. Liu, K. Liu, C.~Y. Cui, and M.~Q. Huang, \href{https://doi.org/10.1088/1361-6471/aa68f1}{J. Phys. G {\bf 44}, 075006 (2017)}.

\end{thebibliography}
\end{document}